\documentclass[sigconf]{acmart}
\usepackage{enumitem}
\AtBeginDocument{%
  }

\setcopyright{acmcopyright}
\copyrightyear{2025}
\acmYear{2025}
\begin{document}

\title{WeVibe: Weight Change Estimation Through Audio-Induced Shelf Vibrations In Autonomous Stores}

\author{\large Jiale Zhang}
\orcid{0000-0003-0688-564X}
\email{jiale@umich.edu}
\affiliation{%
  \institution{University of Michigan}
  \streetaddress{1301 Beal Ave}
  \city{Ann Arbor}
  \state{MI}
  \country{USA}
  \postcode{48109-2122}
}

\author{\large Yuyan Wu}
\orcid{0009-0009-3152-939X}
\email{wuyuyan@stanford.edu}
\affiliation{%
  \institution{Stanford University}
  \streetaddress{}
  \city{Stanford}
  \state{CA}
  \country{USA}
  \postcode{94305}
}

\author{\large Jesse R Codling}
\orcid{0000-0001-8355-7186}
\email{codling@umich.edu}
\affiliation{%
    \institution{University of Michigan}
    \city{Ann Arbor}
    \state{Michigan}
    \country{USA}
}

\author{\large Yen Cheng Chang}
\orcid{0009-0007-1986-9485}
\email{yencheng@umich.edu}
\affiliation{%
    \institution{University of Michigan}
    \city{Ann Arbor}
    \state{Michigan}
    \country{USA}
}

\author{\large Julia Gersey}
\email{gersey@umich.edu}
\affiliation{%
  \institution{University of Michigan}
  \city{Ann Arbor}
  \state{MI}
  \country{USA}}

\author{\large Pei Zhang}
\orcid{0000-0002-8512-1615}
\email{peizhang@umich.edu}
\affiliation{%
  \institution{University of Michigan}
  \streetaddress{1301 Beal Ave}
  \city{Ann Arbor}
  \state{MI}
  \country{USA}
  \postcode{48109-2122}
}

\author{\large Hae Young Noh}
\orcid{0000-0002-7998-3657}
\email{noh@stanford.edu}
\affiliation{%
  \institution{Stanford University}
  \streetaddress{}
  \city{Stanford}
  \state{CA}
  \country{USA}
  \postcode{94305}
}

\author{\large Yiwen Dong}
\orcid{0000-0002-7877-1783}
\email{ywdong@stanford.edu}
\affiliation{%
  \institution{Stanford University}
  \streetaddress{}
  \city{Stanford}
  \state{CA}
  \country{USA}
  \postcode{94305}
}

\renewcommand{\shortauthors}{Jiale et al.}

\begin{abstract}
Weight change estimation is crucial in various applications, particularly for detecting pick-up and put-back actions when people interact with the shelf while shopping in autonomous stores. Moreover, accurate weight change estimation allows autonomous stores to automatically identify items being picked up or put back, ensuring precise cost estimation. However, the conventional approach of estimating weight changes requires specialized weight-sensing shelves, which are densely deployed weight scales, incurring intensive sensor consumption and high costs. Prior works explored the vibration-based weight sensing method, but they failed when the location of weight change varies.

In response to these limitations, we made the following contributions: (1) We propose WeVibe, a first item weight change estimation system through active shelf vibration sensing. The main intuition of the system is that the weight placed on the shelf influences the dynamic vibration response of the shelf, thus altering the shelf vibration patterns. (2) We model a physics-informed relationship between the shelf vibration response and item weight across multiple locations on the shelf based on structural dynamics theory. This relationship is linear and allows easy training of a weight estimation model at a new location without heavy data collection. (3) We evaluate our system on a gondola shelf organized as the real-store settings. WeVibe achieved a mean absolute error down to 38.07g and a standard deviation of 31.2g with one sensor and 10\% samples from three weight classes on estimating weight change from 0g to 450g, which can be leveraged for differentiating items with more than 100g differences.
\end{abstract}


\begin{teaserfigure}
\setlength{\abovecaptionskip}{10pt}
    \centering
    \includegraphics[width=\linewidth]{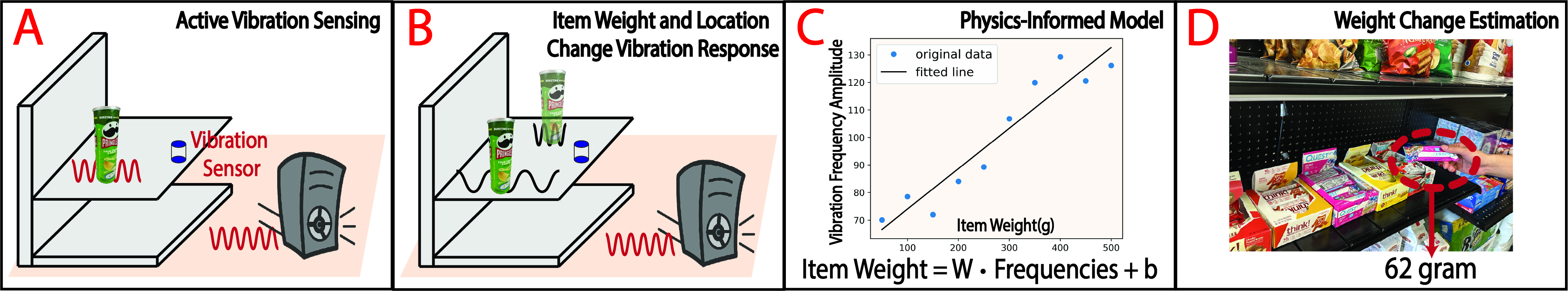}
    \caption{(A) WeVibe leverages the audio-induced vibration propagating to the shelf to create an active vibration sensing environment, reducing the need for sensors attached to the shelf. (B) The different item weights and locations impact the original shelf vibration response differently, reflected by the vibration signal collected by the sensor. (C) According to structural dynamics, a physics-informed relationship (a linear model) between the shelf vibration response and item weight across different locations is characterized to estimate the item weight change on the shelf with improved data efficacy. (D) Through physics-informed learning, WeVibe can estimate the weight change quickly at a new location with fewer sensors and data.}
    \Description{figure description}
    \label{fig:overview}
\end{teaserfigure}

\maketitle

\section{Introduction}
\label{sec:Section1}
Item weight change estimation is crucial in many applications~\cite{zhang2020vibroscale,he2013food,bonde2021pignet}, especially in the context of autonomous retail environments~\cite{ruiz2019aim3s,falcao2020faim,rohal2024don}. Autonomous stores, equipped with technologies to identify which items customers have taken, facilitate automatic checkouts without the need for cashiers. This process typically relies on camera systems; however, cameras can fail when their view is blocked. Consequently, integrating precise weight estimation can significantly enhance the robustness of item detection and ensure accurate cost assessments, thereby overcoming limitations posed by visual-only systems.

\begin{figure*}[t]
    \centering
    \includegraphics[width=0.8\linewidth]{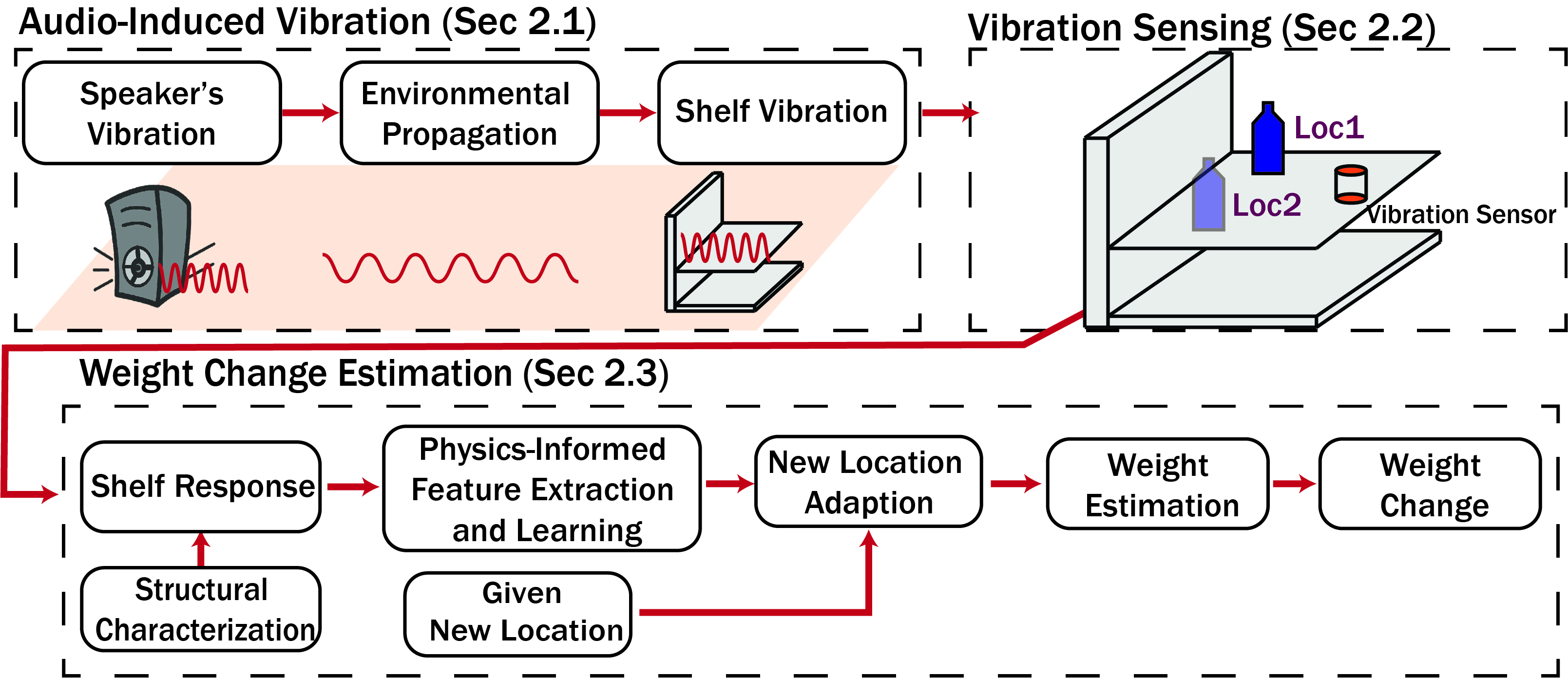}
    \caption{WeVibe system diagram comprises (a) An audio-induced vibration module designed for the active vibration sensing environment. (b) The vibration sensing module that takes shelf vibration response when there are different weights on the shelf. (c) The weight change estimation module that leverages physics-informed knowledge to extract features and develop a learning model for weight estimation. The weight change is achieved by calculating the difference between the two weight estimation results.}
    \label{fig:Figure2}
\end{figure*}

However, the conventional implementation of weight change estimation in autonomous stores incurs significant sensor costs. The recent development of autonomous stores employs smart pressure-based weight sensing shelves to achieve automatic item detection~\cite{ruiz2019aim3s,falcao2021isacs}, which involves densely installed weight scales underneath the items. Even though these weight-sensing shelves provide accurate estimation, each shelf must be equipped with 12 to 24 weight sensors to monitor changes comprehensively~\cite{ruiz2019aim3s,gu2022tracking} and achieve precise analysis of item weights at various locations on the shelf, heavily elevating the sensor cost. Furthermore, considering more gondolas in the stores, densely installed weight scales might lead to a higher total cost and pose scalability challenges.

Vibration sensing has shown great promise in estimating the weight, but the unknown relationship between vibration and item weight on the shelf prevents it from being used in autonomous stores. Many prior works indicate that the change of weight results in different structural responses~\cite{sekiya2018simplified,codling2021masshog,mirshekari2016characterizing, mirshekari2021obstruction}, inspiring us to explore the capability of vibration on estimating item weight change on the gondola shelf. Nevertheless, these works focus more on modeling the relationship between the single item staying at the same location and the structure vibration response. It cannot be adopted in complex scenarios like autonomous stores because the weight change may happen at different locations on the shelf, and there are other items that stay together. It is also impractical to collect a large amount of data to model the relationship between shelf vibration response, item weight, and item location because the variability introduced by different item weights and their locations leads to a vast number of possible data scenarios.

To tackle these problems, we make the following three contributions:
\begin{enumerate}[label=(\arabic*)]
  \item We propose WeVibe: The first system that utilizes audio-induced vibrations from a speaker to detect weight changes on the shelf during shopping using one vibration sensor at best.
  \item We model a structure-dynamics-informed relationship between the shelf vibration response and item weight across multiple locations on the shelf, allowing an easy training of the weight estimation model at a new location without heavy data collection.
  \item  We validate our system with a real-world shopping layout, demonstrating the efficacy of WeVibe in real-world scenarios.
\end{enumerate}
WeVibe adopts active vibration sensing at a lower sensor cost. Since the items on the shelf cannot generate vibration by themselves, WeVibe takes a speaker as an exterior vibration source and estimates item weight change by analyzing the shelf vibration signal. The vibration wave from the speaker can propagate to the shelf through environmental structures such as the floor and gondola, creating a vibration pattern across the entire shelf. Therefore, one vibration sensor can capture the difference of the whole shelf vibration response induced by the weight change, reducing the number of sensors.

On the other hand, WeVibe features a physics-informed relationship (linear model) between the shelf vibration response and item weight across different item locations, improving the data efficacy. Through empirical observation, we discover that the increasing or decreasing item weight leads to the same rising or falling trend on some frequencies of impulse vibration propagated on the shelf, resulting in an assumption of linearity. Then, we justify this linearity through the theoretical derivation, developing this structure-dynamics-informed relationship between the vibration response and item weight as a linear model. This physics-informed knowledge solidifies the linear assumption during the learning process, allowing WeVibe to adapt to a new location by training a weight estimation model easily for each location of interest with two or three weight classes, alleviating the data collection effort compared with the deep learning method without prior knowledge.

The rest of the paper is organized as follows: Section ~\ref{sec:System Overview} provides a system overview of WeVibe, illustrating the active vibration sensing setup for less sensor and how WeVibe processes the weight change estimation. Section ~\ref{sec:Structure-dynamics-informed modeling} explains the characterization of the structure-dynamics-informed relationship for less data collection, followed by the evaluation in Section ~\ref{sec:System Evaluation}. Section ~\ref{sec:Related Work} goes through an overview of the autonomous store and then introduces the related work of weight sensing. Finally, we conclude our work in Section ~\ref{sec:Conlcusion}.

\section{System Overview}
\label{sec:System Overview}
This section discusses the WeVibe system overview using active vibration sensing and structure-dynamics-informed modeling for item weight change estimation. WeVibe constructs an active vibration sensing environment through the audio-induced shelf vibration (Section~\ref{sec:audio induced vibration}), reducing the sensor cost. By playing sound from a speaker, a mechanical vibration wave is generated. This vibration will propagate through the structure, like the floor, and arrive at the gondola shelf. The vibration sensing module then captures the shelf vibration response (Section~\ref{sec: vibration sensing}). When the item weight changes, it will impact the original shelf vibration differently, which can be leveraged to estimate the item weight. Through the knowledge of structural dynamics, WeVibe characterizes these different shelf responses and develops the physics-informed features and machine-learning model to adapt to a new item location for weight estimation quickly. Finally, the difference between the two weight estimations is taken as the weight change estimation result (Section~\ref{sec: Weight Change Estimation}).

\subsection{Audio-Induced Vibration}
\label{sec:audio induced vibration}
WeVibe creates an active vibration sensing environment by playing sound from a speaker next to the gondola. When the speaker plays sound, the interior components like the cone and voice coil generate mechanical vibration. These vibration waves can propagate through the environment by exciting the structure particle movement. There are two basic wave types: Longitudinal wave and transverse wave~\cite{noh2023dynamics,yuan2022spatial}. The particles moving parallel to the wave propagation form the longitudinal wave. The particles moving perpendicular to the wave propagation form the transverse wave. With these two wave types, the mechanical vibration from the speaker can travel to the shelf.

A periodic impulse is selected to burst out a strong vibration wave capable of reaching the shelf. The shelf response resulting from each impulse is taken as one sample for weight estimation. Many types of sound can be potentially adopted for developing the vibration wave, such as constant tone, frequency sweep, and impulse. To optimize the performance of WeVibe, a wider frequency band and a more vigorous signal intensity are desired. We employ an impulse signal. The impulse signal bursts out intensively in a very brief period, so it can provide a sudden and forceful push to the speaker's components, causing a strong vibration. Additionally, the impulse sound has a broad frequency spectrum, preparing a vast feature pool for further signal processing. The speaker's volume is set as an average person's speaking volume while keeping enough intensity that the vibration sensor can get the signal.

\subsection{Vibration Sensing}
\label{sec: vibration sensing}
\begin{figure}
    \centering
    \includegraphics[width=\linewidth]{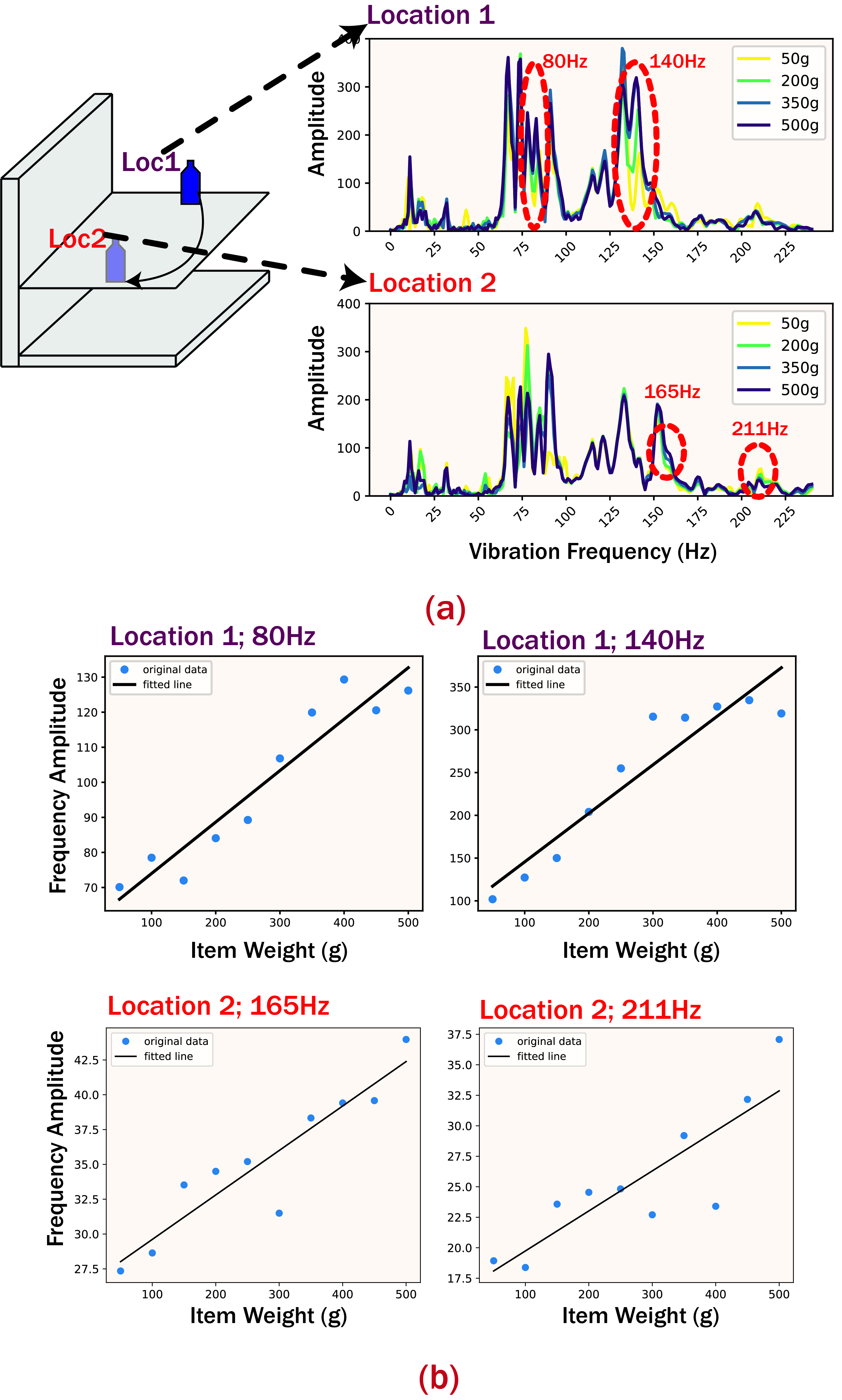}
    \caption{(a) The plot shows the frequency spectrums of different weights of a single item at two locations. For the same location, some weight-sensitive frequencies increase or decrease while the weight of the item increases, as indicated by the red circle. Furthermore, when the item changes location, the overall frequency spectrum has a more significant change, and the weight-sensitive frequencies also shift. (b) We Further plot the result of linear regression on the highlighted frequencies and item weight at both locations. Even though some points deviate from the fitted line, the visualization gives rise to the assumption of linearity.}
    \label{fig:frequency difference}
\end{figure}

The weight of an item placed on a surface influences the vibration signal (as shown in Figure~\ref{fig:frequency difference}) due to the changes in the structural properties (e.g., mass, stiffness, and damping ) after an item is added or removed. When a heavier item is placed on a surface, it tends to absorb and dampen the vibrations more significantly, leading to variations in signal amplitude, frequency response, and decay rate compared to lighter items. Conversely, a lighter item has a weaker impact on the vibration signal with different characteristics. These distinctions arise because the weight affects the surface's structural properties. By accurately capturing and analyzing these variations in vibration signals, it is possible to determine the item's weight with high precision.

On the other hand, the location of where the weight change happens also affects the shelf vibration response. Through a physics-informed characterization of the shelf vibration response, item weight, and item location, WeVibe quickly provides each location with a dedicated learning model, which will be explained in Section ~\ref{sec:Structure-dynamics-informed modeling}.

Adopting the active vibration sensing method allows fewer sensors to be attached to the shelf than the smart weight sensing shelf. This is because the vibration signal captured by one single sensor can still effectively represent the structural characteristics of the entire shelf, which we will discuss more in Section ~\ref{sec:Structure-dynamics-informed modeling}. Studies have shown that these single-point vibration signals are helpful in detecting structural anomalies and assessing the integrity of various structures\cite{sekiya2018simplified,obrien2020using,yu2016state}. This insight allows WeVibe to employ one vibration sensor to capture the shelf's vibrational response and indicate weight information through further signal processing. Multiple sensors can also be attached to the shelf to improve the system's robustness and accuracy.

\subsection{Weight Change Estimation}
\label{sec: Weight Change Estimation}
We leverage the knowledge of structural dynamics to characterize a physical model between the shelf vibration response and item weight to reduce the need for data collection. Through empirical studies, we first notice that vibration frequencies differ when the item weight changes. Furthermore, some frequency amplitudes show the same increasing or decreasing trend while item weight increases or decreases. Therefore, we assume a linear model between the vibration frequency spectrum and item weight. This assumption is then validated through the theoretical derivation based on the structural dynamics, detailed in section~\ref{sec:Structure-dynamics-informed modeling}. Therefore, WeVibe applies the physics-informed feature extraction and learning model to the shelf responses to estimate the item's weight. Given a new location, Wevibe can exploit this physics-informed relationship to quickly train a weight estimation model with two weight classes at best and apply the estimation to a broader range of weight. Finally, the difference between the two weight estimations is taken as the weight change estimation result.

\section{Structure-Dynamics-Informed Vibration Modeling For Reduced Data Need}
\label{sec:Structure-dynamics-informed modeling}

\begin{figure*}[t]
    \centering
    \includegraphics[width=0.8\linewidth]{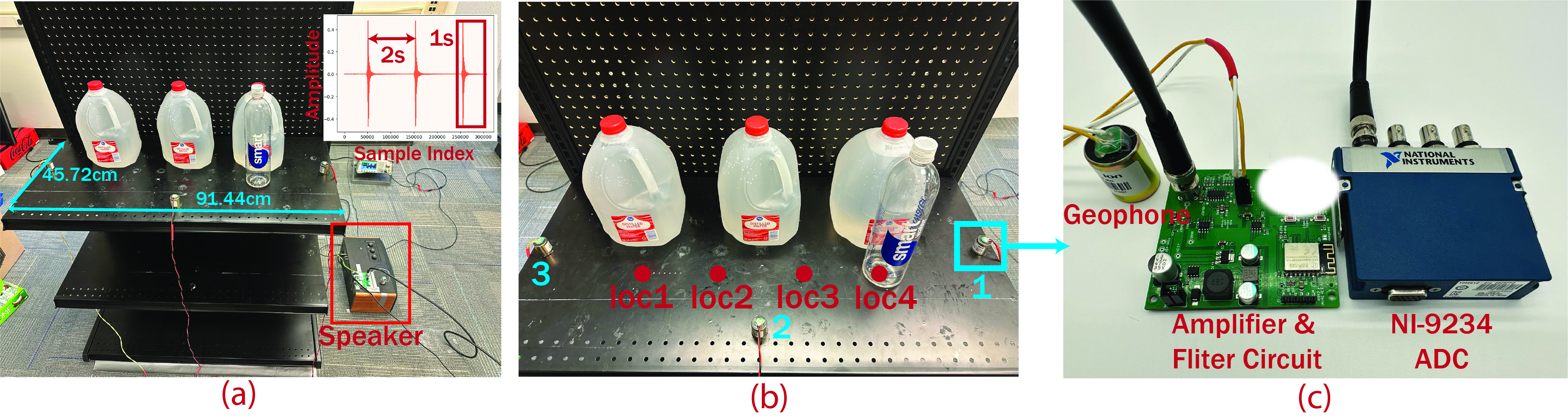}
    \caption{(a) shows the store gondola and active vibration sensing setup with a speaker next to the gondola. The top right signal clip shows an example of our given vibration signal. (b) gives a more detailed view of the item location and sensor location. The different weight classes are taken by changing the amount of water in the 1L water bottle. (c) provides an overview of our vibration sensing module.}
    \label{fig:Experiment Setup}
\end{figure*}

WeVibe leverages structural dynamics knowledge to characterize the physics-informed relationship between shelf vibration and item weight change. It is observed that the item weight change and the vibration frequency spectrum follow the same increasing or decreasing trend, as indicated in Figure~\ref{fig:frequency difference}. This observation inspires a linear relationship assumption, which is justified by the derivation according to the structural dynamics. In addition, WeVibe can quickly adapt to multiple locations with several weight classes based on the observation that the linearity is kept at various locations, though the coefficients are different.

Section~\ref{sec:empirical study} will first illustrate our empirical observations, leading us to the assumption of linearity across locations. Then, Section~\ref{sec:theory} validates our observations through the theoretical model of the interaction between the vibration frequency spectrum, item weight, and item location. It points out the hidden linearity in the theoretical formulation. Section~\ref{sec: Feature Extraction and Learning} provides more detail on physics-informed feature extraction.

\subsection{Empirical Study On The Relationship Between Shelf Vibration and Item Weight}
\label{sec:empirical study}
As Figure~\ref{fig:frequency difference} (a) shows, the vibration frequency spectrums differ when various item weights and locations are presented. For the same location, the vibration frequency spectrum shows minor differences between the four weights. However, as highlighted by the red circle, these frequencies tend to increase or decrease as the weight increases monotonically. Therefore, the linear regression is applied to these frequencies and more weights to visualize their correlations, as shown in Figure~\ref{fig:frequency difference} (b). Even though some points deviate from the fitted line, the result suggests a possible linearity between the weight and these mentioned frequencies, which might exist at both locations. 

On the other hand, the various item locations lead to a more distinct change in the vibration frequency spectrums compared with the item weight. When the item location changes, the weight-sensitive frequencies shift to somewhere else, and the overall frequency spectrum envelope also significantly changes. These factors result in the different linear relationships shown in Figure~\ref{fig:frequency difference} (b). With these observations, we assumed that the vibration frequency spectrum is linearly related to the item weight across multiple locations on the shelf under our active vibration sensing environment.

\subsection{Theoretical Derivation On The Relationship Between Shelf Vibration and Item Weight}
\label{sec:theory}
\begin{figure}[tbh]
    \centering
    \includegraphics[width=\linewidth]{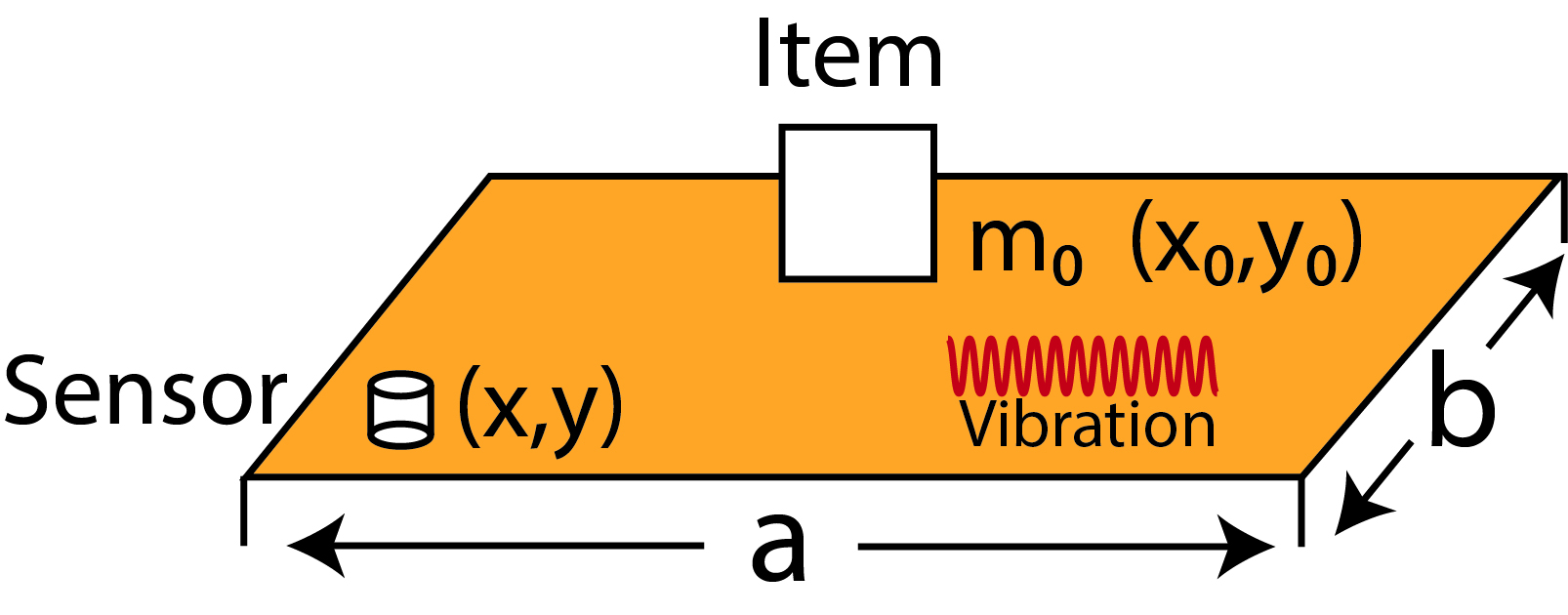}
    \caption{The simplified shelf model for developing the theoretical model.}
    \label{fig:Theory}
\end{figure}

Using the structural dynamics theory, we characterize the relationship between shelf vibrations and item weight placed on the shelf to validate our assumption. In our analysis, the shelf is modeled as a thin and homogeneous plate with the length $a$ and width $b$, assuming that its thickness is much smaller than its length and width (see Figure~\ref{fig:Theory}), which is commonly valid for the steel shelf in the retail stores. The item is modeled as a point load on the shelf at location $(x_0, y_0)$. Other assumptions include the simply supported boundary and ignorable damping effect. The sensor's location is represented by $(x,y)$.

Based on the Kirchhoff–Love plate theory that describes the behavior of thin plates subjected to forces and moments, the governing equation of the plate vibration can be formulated as Equation ~\ref{eq:gov_eq}, ignoring the damping. In this equation, $D, \rho, \nu$ correspondingly represent the flexural rigidity, mass per unit area, and Poisson ratio of the shelf. $f(x, y, t)$ is the excitation source exerted at location $(x, y)$ at time $t$. $w(x,y,t)$ represents the shelf vertical displacement, i.e., the vibration at location $(x, y)$. Solving Equation ~\ref{eq:gov_eq} based on the simply supported boundary and static initial condition, $w(x, y, \omega)$, the Fourier transform of the time-domain vibration signal, is proportional to its spatial Fourier transform coefficient $\bar{W}(m,n,\omega)$ which is linear to the item weight $m_0$, as shown in Equation ~\ref{eq:solution}. $\bar{F}(m,n,\omega)$ is the 2d spatial Fourier transform of the excitation force $f(x, y, \omega)$ at the sensor location $(x,y)$. $\omega_{mn}$ is a function of $m$ and $n$. When the item's weight is much smaller than the mass per unit area of the shelf, we can approximate the equation further with Taylor expansion as shown in ~\ref{eq:taylor expansion}. In this situation, \textbf{the vibration frequency spectrum is linear to the item's weight.}

\begin{equation}
D\nabla^4 w + \rho \frac{\partial^2 w}{\partial t^2} = f(x, y, t) + \delta(x-x_0)\delta(y-y_0)(m_0g + m_0\frac{\partial^2 w}{\partial t^2})
\label{eq:gov_eq}
\end{equation}

\begin{equation}
\begin{gathered}
w(x, y, \omega) \propto  \bar{W}(m,n,\omega) \\
= \frac{\bar{F}(m,n,\omega)}{-\omega^2 \left( \rho + \textcolor{red}{m_0} \sin\left( \frac{m x_0 \pi}{a} \right) \sin\left( \frac{n y_0 \pi}{b} \right) \right) + D \omega^2_{mn}}
\label{eq:solution}
\end{gathered}
\end{equation}

\begin{equation}
\begin{gathered}
\approx \frac{\tilde{F}(m, n, \omega)}{-\omega^2 \rho + D \omega_{mn}^2} \left( 1 + \frac{\omega^2 \sin \left( \frac{m x_0 \pi}{a} \right) \sin \left( \frac{n y_0 \pi}{b} \right)}{-\omega^2 \rho + D \omega_{mn}^2} \textcolor{red}{m_0} \right)
\label{eq:taylor expansion}
\end{gathered}
\end{equation}

The vibration response $w(x,y,\omega)$ not only depends on the item weight $m_0$, but also the item location $(x_0, y_0)$ as shown in Equation ~\ref{eq:taylor expansion}. Although this linearity holds true for different item locations, the linear coefficients are different because they depend on $(x_0, y_0)$. Intuitively, this is because different item locations affect different modes of shelf structure. Thus, the shelf vibration response is affected in various ways. To this end, the theoretical derivation validates (1) The shelf vibration frequencies are linearly correlated with the item weight. (2) This linearity is kept across the gondola shelf.

\subsection{Physics-Informed Feature Extraction and Learning}
\label{sec: Feature Extraction and Learning}

\begin{figure*}[t]
    \centering
    \includegraphics[width=\linewidth]{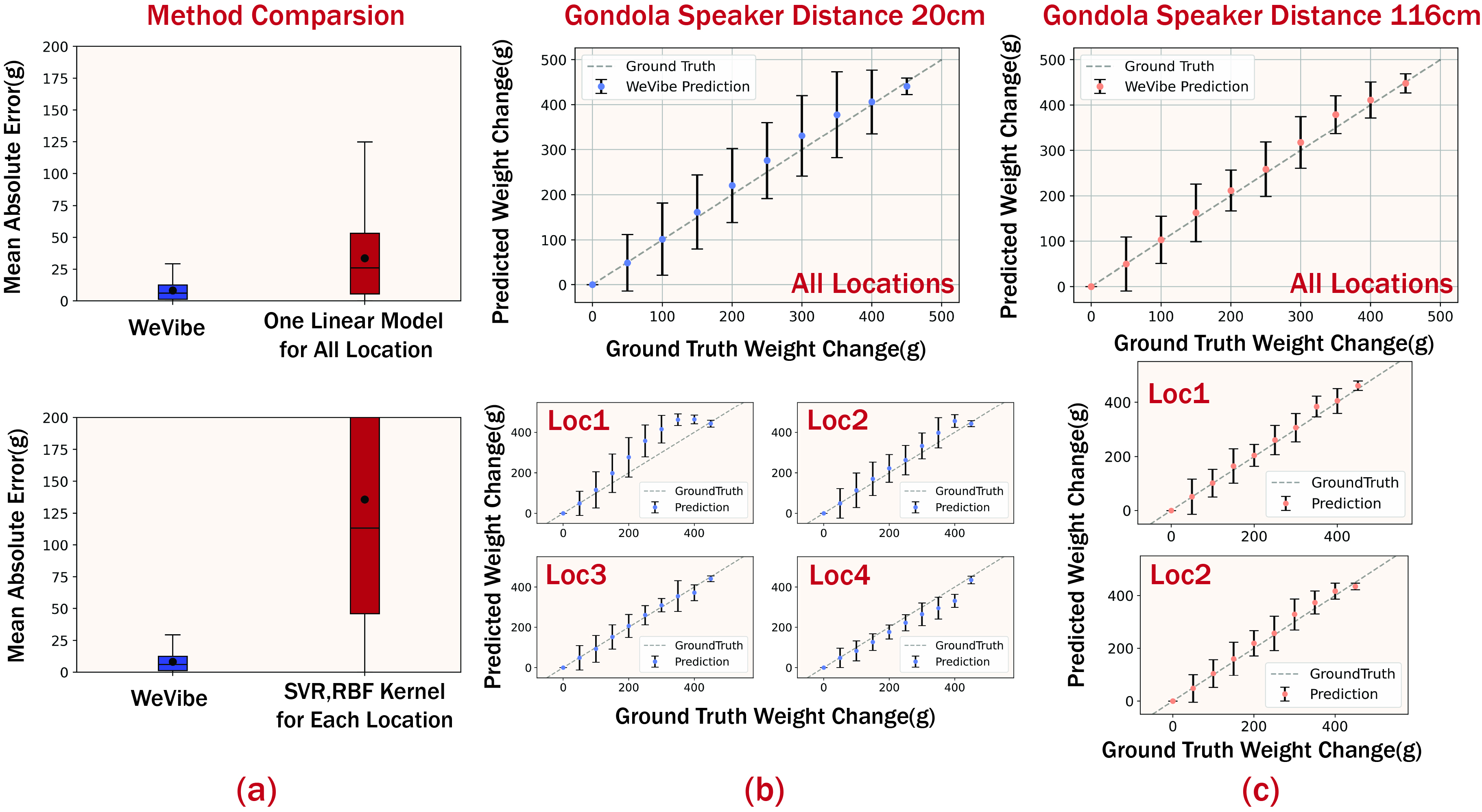}
    \caption{The WeVibe system evaluation. (a) The comparison between WeVibe and the other two methods: Using one linear model for all locations and using the non-linear model for each location while using 10\% of all weight classes in training and one vibration sensor. WeVibe outperforms both approaches with a significant improvement. (b)\&(c) The weight change estimation result of WeVibe, taking 10\% of 3 weight classes in training and one vibration sensor. The result suggests that WeVibe can almost certainly distinguish weight changes bigger than 100g, which can be utilized to detect whether a can of chips or a bottle of water is taken or put back.}
    \label{fig:Evaluation1}
\end{figure*}

With theoretical derivation and empirical study, WeVibe builds a solid physics-informed feature extraction and learning model. WeVibe takes the shelf vibration response from one whole impulse vibration as one sample so that the feature will contain the information of both transient and steady state~\cite{huang2020vibration}. WeVibe then takes the vibration frequency spectrum through the Fourier transform. As Figure~\ref{fig:frequency difference} suggests, the shift of item location leads to the change of weight-sensitive frequencies, which is challenging to know beforehand. Therefore, instead of choosing one or two specific weight-sensitive frequencies, WeVibe takes the full range of vibration frequency spectrum except for the noise-occupied region, 50Hz to 240Hz, as the feature for the learning model.

\section{WeVibe Evaluation}
\label{sec:System Evaluation}
We evaluate WeVibe's performance on a standard gondola with a simple item layout first and then a real-world item layout for weight change estimation from three aspects: ~\ref{eval:System} System, ~\ref{eval:data} The usage of Data, and \ref{eval:sensor} The usage of sensor.

\subsection{Experiment Setup}
\subsubsection{Vibration Sensing Environment Setup}
We evaluate WeVibe on a real gondola shelf that is commonly used in the real-world store (Figure~\ref{fig:Experiment Setup} (a)). The shelf is manufactured from MFired Store Fixtures\cite{mfried} with steel and mounted on a double-sided gondola. The shelf length is 91.44cm, and the width is 46.72cm. We keep the other shelves empty except for the one used for evaluation. We put one 1L water bottle on the shelf and estimate the weight change of water in the 1L bottle at four different locations as indicated in Figure~\ref{fig:Experiment Setup} (b). We also put an extra three 3.78L water bottles on the shelf because the whole shelf is rarely empty in real-world cases.

An Edifier D12 speaker plays an impulse signal composed of a short sinc function from a function generator to create a vibration. The sinc function has a 10Hz central frequency and 15V peak-to-peak voltage. It bursts once every 2s, as shown in the top right corner of Figure~\ref{fig:Experiment Setup} (a). The speaker is 20cm from the bottom right edge of the gondola and faces to the right. Additionally, to evaluate whether WeVibe can work on a gondola with a greater distance from the speaker, we move the speaker further away to 116cm, increasing the distance as the length of the shelf. 

\subsubsection{Vibration Sensing Module}
The vibration signal is captured by the vibration sensing module composed of (1) A geophone, (2) An amplifier \& filter circuit, and (3) An NI-9234\cite{ni9234} analog-to-digital converter (ADC) as shown in Figure~\ref{fig:Experiment Setup} (c). The SM-24 geophone is employed to collect vibration signals~\cite{SM_24} and has an enhanced frequency response between 10Hz and 240Hz. It has been shown with prospective performance on collecting structural vibration\cite{pan2017footprintid,bonde2021pignet,mirshekari2021obstruction,zhang2023vibration}. The geophone can measure vibrations by converting the velocity of the mechanical movement of the structure into an electrical signal that can be quantitatively analyzed. To increase the signal-to-noise ratio and remove the background noise, we design a programmable amplifier \& filter circuit. We set the gain to 35 and removed the frequency component outside SM-24 geophone functional bandwidth. Finally, the NI ADC converts the analog signal into a 24-bit digitalized signal with a sampling rate of 51.2kHz per second. The NI ADC also allows a synchronized collection of multiple-channel vibration signals, which prepares us to evaluate different sensor configurations (Section~\ref{eval:sensor}).

There are four sensing modules in total; one is on the speaker to monitor the input signal and is considered as the reference signal. The other three are placed in the middle of the right shelf edge, the middle of the front shelf edge, and the middle of the left shelf edge, and they are used to evaluate the performance of sensors at different configurations.

\subsubsection{Data Collection and Pre-Processing}
We collected ten weight classes by changing the amount of water in a 1L water bottle, ranging from 50g to 500g, with 50g spacing at four different locations with a 15.24cm interval, as shown in Figure ~\ref{fig:Experiment Setup} (b). The ground truth weight is the same as the weight of the 1L water bottle. We ignore the weight of the 3.78L water bottle because their weights are not changing. Each weight has 28 samples at each location. There are 1120 samples in total. The collected weight classes result in the absolute weight change from 0g to 450g with 50g spacing. Furthermore, we repeat the data collection on item locations 1 and 2 with a further distance(116cm) between the speaker and gondola to investigate whether WeVibe can work for a gondola at a new location.

After data collection, we pre-process the data into individual samples. All the vibration signals are first aligned with the signal collected on the top of the speaker for synchronization. Then, as shown in Figure~\ref{fig:Experiment Setup} (a), each sample begins 0.1s before the beginning of the impulse signal and lasts 1s so that it can fully capture the shelf responses in the excitation stage and steady stage\cite{huang2020vibration}. Then, we apply the Fourier transform to get the frequency spectrum and use frequencies from 50Hz to 240Hz with 1Hz spacing as the feature vector~\ref{eq:feature vector}. To further reduce the dimension of the feature vector and improve learning efficacy, Principle Components Analysis (PCA) ~\cite{mackiewicz1993principal} is then applied for every feature vector to extract the most valuable information. Finally, the features from the same location are fed into a linear regression model with L2 regularization for learning the weight information.

\begin{equation}
\begin{gathered}
Feature Vector = [f_{50Hz},f_{51Hz},...,f_{240Hz}] = [f_1,f_2,...,f_{191}]
\label{eq:feature vector}
\end{gathered}
\end{equation}

\subsection{Evaluation I: WeVibe Overall Performance}
\label{eval:System}
Figure~\ref{fig:Evaluation1} shows the result of WeVibe system evaluation. Figure~\ref{fig:Evaluation1} (a) indicates an ablation test of WeVibe methodology, additionally validating the rationality of assigning a linear model for each location. Figure~\ref{fig:Evaluation1} (b) and (c) show the weight change estimation result with one standard deviation range. After estimating the weight, the difference between any two weights is taken as the final result. Since the negative and positive weight differences are symmetric, only the positive weight differences are shown in the following figures.

We first compare WeVibe with the other algorithms as shown in Figure~\ref{fig:Evaluation1} (a). Taking 10\% of training samples (three samples) of ten weight classes and one sensor, WeVibe mean absolute error outperforms the method that uses one linear model for all locations with 4.2X improvement, proving that different linear relationships are kept at various locations. On the other hand, WeVibe's means absolute error outperforms the method that uses a non-linear model for each location with 17X improvement, indicating the correctness of the linear relationship. 

Taking 10\% of training samples of three weight classes (50g, 300g, and 500g) and one sensor, Figure~\ref{fig:Evaluation1} (b) and (c) suggest WeVibe performances for gondolas with different distances with speakers. When the distance is 20cm, WeVibe achieves an overall mean absolute error of 48.15g with a standard deviation of 44.96g. When the distance is 116cm, the overall mean absolute error is 38.07g, and the standard deviation is 31.2g. These results suggest that WeVibe can correctly differentiate weight change bigger than 100g in most cases, which only takes three weight classes at every location and one vibration sensor. It is accurate enough to be leveraged for detecting whether items like a can of chips or a bottle of water are taken or put back. For lightweight items such as a bar of chocolate, a fine granularity of training weight class might be necessary. A minimum of sensors attached to the surface and better data efficacy are also suggested, which will be detailed in the following sections. Furthermore, both distances achieve a similar result, indicating a better ubiquity because the linearity between shelf vibration response and item weight still exists when the gondola is at a different location. 

\subsection{Evaluation II: The Amount of Required Data}
\label{eval:data}

\begin{figure}[tbh]
    \centering
    \includegraphics[width=0.6\linewidth]{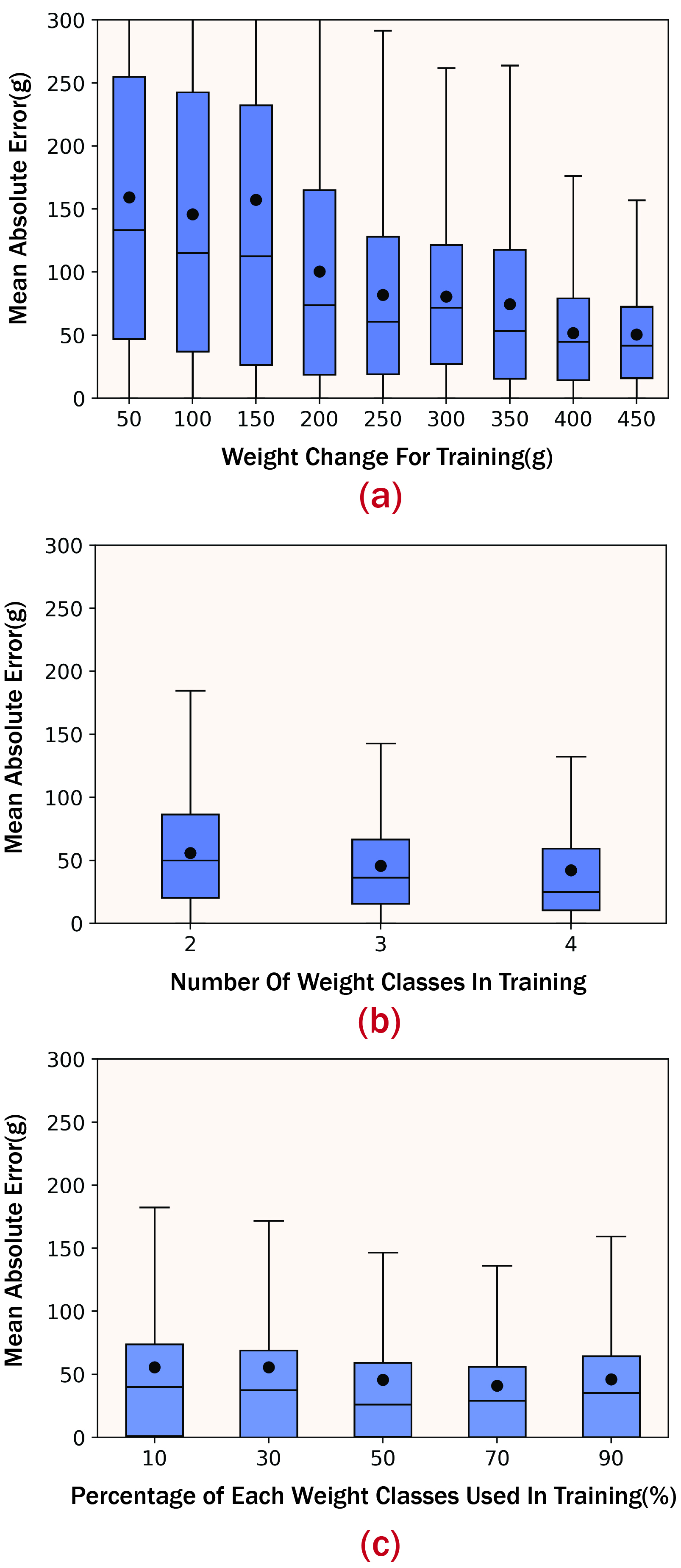}
    \caption{(a) With two training weight classes, the smaller one is 50g, and the bigger one gradually increases. It can be seen that the error gets smaller with a bigger range. (b) Adding more training weight classes decreases the median and mean MAE, suggesting a tradeoff between data collection effort and performance. The improvement from two to three weight classes is more significant than the improvement from three to four, leading us to select three weight classes for evaluation.  (c) The increasing amount of training data per weight class gives comparable performances. Therefore, three samples are selected in our case for better data efficacy.}
    \label{fig:Evaluation2}
\end{figure}

Three evaluations are conducted to validate the improved data efficacy based on the physics-informed model. To evaluate the tradeoff between the performance and the amount of training data, we start from two weight classes (one weight change) and find that when the minimum and maximum weights are included in the training set, the performance reaches the best. The minimum weight 50g is kept in the training set and the bigger weight increases gradually (e.g., 50g and 100g, 50g and 150g,\ldots, 50g and 500g). As weight change gets bigger, the performance gets better. The inclusion of minimum weight and maximum weight reaches the best, as shown in Figure~\ref{fig:Evaluation2} (a). It reflects that when the training data includes the full range of weight information, it can better capture the weight falling within this range, which is very similar to the property of a linear model.

Then, the number of training weight classes increases while keeping the minimum and maximum weight in the training set, evaluating the tradeoff between performance and the number of weight classes. The number of weight classes increases by interpolation: 50g and 500g; 50g, 300g and 500g; 50g, 200g, 350g, and 500g. Figure~\ref{fig:Evaluation2} (b) suggests that the increasing number of training weight classes provides an improvement. However, the improvement is more evident from two to three weight classes. The performance of three weight classes and four weight classes are very close. In the actual application, the number of necessarily collected weight classes significantly depends on the possible slightest weight change between items. In this case, three training weight classes are employed because there is no significant improvement in selecting four training weight classes.

Figure~\ref{fig:Evaluation2} (c) suggests that 10\% training data for each weight class is enough. When keeping three weight classes selected from the previous evaluation, we investigate the performance while increasing the amount of training data per weight class. It turns out that the increasing amount of training data per weight class gives comparable performances. It is probably because most data samples within the same weight class are similar. Referring back to Section~\ref{sec:theory}, the vibration frequency spectrum should be much the same as long as the item weight and item location do not change. Therefore, we select 10\% training data to minimize the data collection effort.

\subsection{Evaluation III: The Amount of Required Sensor}
\label{eval:sensor}
\begin{figure}[t]
    \centering
    \includegraphics[width=0.8\linewidth]{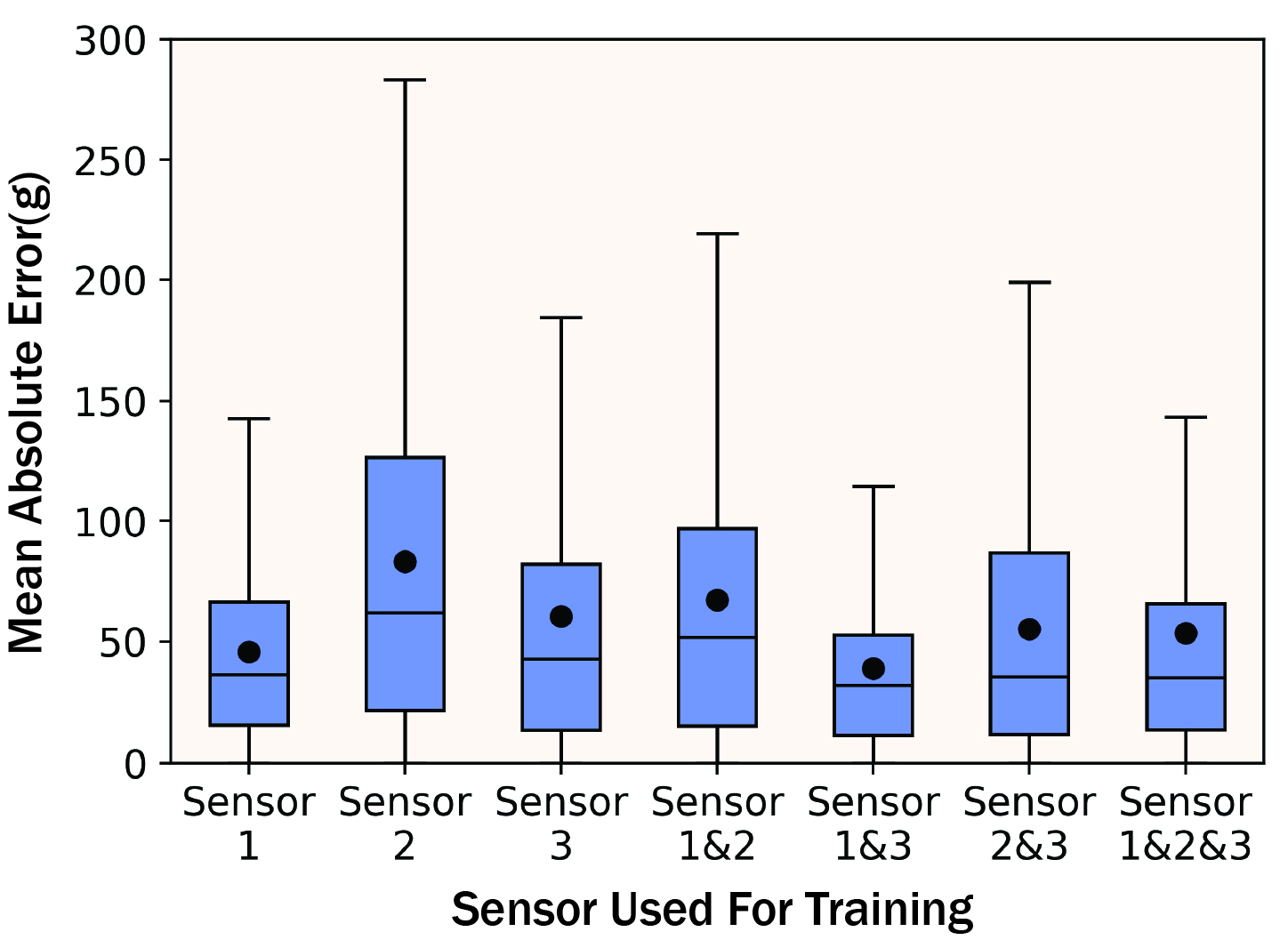}
    \caption{This figure illustrates the tradeoff between weight change estimation and different sensor configurations used during training. Sensor 1\&3 reaches the best performance, but sensor 1 and sensor1\&2\&3 also have a close performance. Given no prior knowledge of which sensor can perform the best, it is reasonable to apply all three sensors. However, if we know the best sensor, it should be used to reduce the sensor cost.}
    \label{fig:Evaluation3}
\end{figure}

Figure~\ref{fig:Evaluation3} shows the performance comparison with different sensor configurations in the training process, suggesting that the combination of three sensors is comparable to a single sensor with the best performance. Figure~\ref{fig:Experiment Setup} (b) shows the location of three different sensors. Figure~\ref{fig:Evaluation3} shows that the performance of sensor 1 is the best out of three single-sensor usage cases, and it is very close to the best performance: sensor 1\&3. This observation leads us to employ sensor 1 to minimize the sensor usage. On the other hand, it is also observed that the combination of three sensors is also very close to the best performance. It results in a compromise of using all three sensors if we don't know which one can give the best performance, which is the case of our real item-layout evaluation.

\subsection{Evaluation IV: Real Item-Layout Evaluation}

\begin{figure}[tbh]
    \centering
    \includegraphics[width=\linewidth]{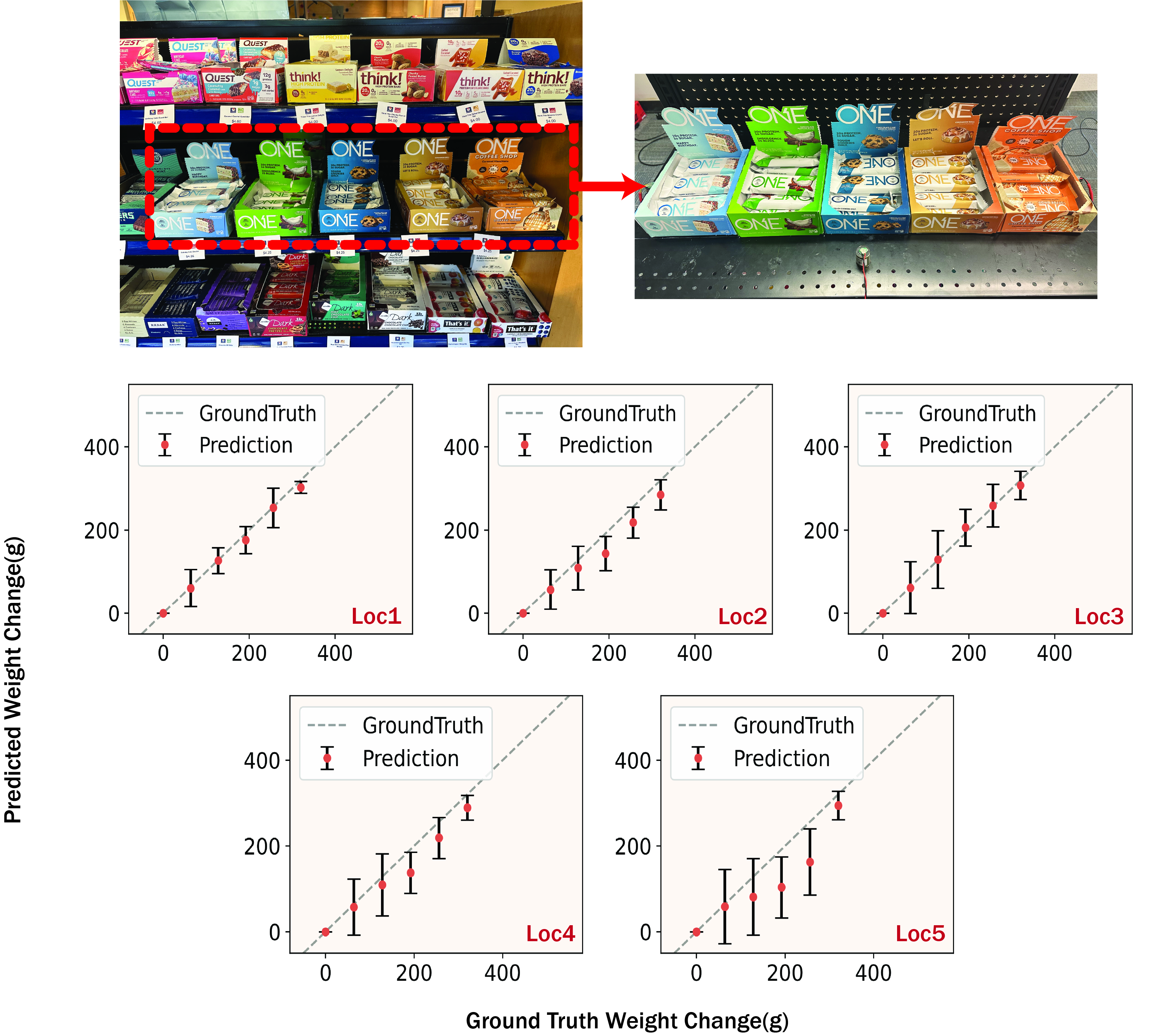}
    \caption{WeVibe is evaluated on a real item-layout shelf. We bought a shelf of protein bars and placed them under the WeVibe sensing environment. With a mean absolute error of 41.05g and a standard deviation of 42.2g, it is expected to detect the case that two or more protein bars are taken or put back according to the weight change estimation.}
    \label{fig:Evaluation Real}
\end{figure}

As shown in Figure~\ref{fig:Evaluation Real}, WeVibe is evaluated with a real item-layout. We bought a shelf of protein bars and placed them on our shelf with the same organization in the store. Six weight samples are collected for each box: Full box, one protein bar is taken,\ldots, and five protein bars are taken. Each protein bar ranges from 59g to 64g, so an average weight of 61g is assigned for each protein bar. The label is the weight of the total shelf load, ranging from 2997g to 3302g (the sum of five boxes). The training set comprises 10\% samples of three weight classes (2997g, 3180g, and 3302g). Three sensors are used together because we assume no prior knowledge of the best sensor is provided.

WeVibe achieves a mean absolute error of 41.05g and a standard deviation of 42.2g for all locations. It suggests that WeVibe can successfully identify these cases based on the weight change estimation result when two or more protein bars are taken or returned. For each location, WeVibe shows various performances. Location 1 gives the best result with 23.34g mean absolute error and 28.3g standard deviation, which likely works for one protein bar. Location 5 shows a more considerable standard deviation, which might not be functional for the lightweight items in the actual store. It might be due to the shelf's elastic deformation. The amount of elastic deformation of the shelf surface might bring extra variation on the shelf vibration response, leading to more data collection at some specific locations. At location 5, more weight classes are required to provide better estimation because there might be a small amount of elastic deformation happening at these locations when the item weight changes, leading to more variances in the final weight change estimation.

\section{Related Work}
\label{sec:Related Work} 
A range of sensing technologies are employed in autonomous stores for item detection. We examine these works and then focus on weight estimation through vibration sensing, which inspired WeVibe.

\subsection{Item Detection In Autonomous Store}
\subsubsection{Vision-only.} The most prevalent sensing approach for item detection involves extensive camera deployment. This approach employs deep-learning object recognition models \cite{he2017mask,wang2023yolov7,tan2020efficientdet} to identify what items customers take. Besides direct prediction, works like \cite{9200182,torres2018text} propose further incorporating the text information on the item to augment the detection accuracy. However, the vision-only approach has three main drawbacks. First, occlusions in the camera line-of-sight hinder accurate item detection. Second, comprehensive store coverage demands many cameras, leading to substantial computational resource consumption. Finally, developing a precise and prompt deep-learning model demands a large and robust dataset, making the training process labor-intensive.

\subsubsection{Vision with other sensing modalities} Introducing other modalities makes the item detection pipeline more robust. For example, \cite{roussos2006enabling,zhang2016mobile} employed RFID with camera. By tagging each item, the accuracy is significantly improved. However, the consequent tagging effort on every item and the incremental cost of each RFID tag quickly make this approach impractical for widespread adoption. Weight sensing in store is pioneered in Aim3s \cite{ruiz2019aim3s}, followed by works with further robust fusion algorithms with the camera system\cite{falcao2021isacs,falcao2020faim}. This method relies on specialized weight-sensing plates capable of discerning weight changes at column-wise locations. However, the necessity for a complete replacement of conventional shelf plates for these specialized units is prohibitive. Additionally, the vulnerability of these plates to wear and malfunction poses further challenges.

\subsection{Weight Estimation Using Vibration}
Weight estimation based on vibration has been scrutinized in structural dynamics. Various theoretical models for analyzing the characteristic behaviors of the beam, like mass distribution and crack location through the vibration frequency spectrum, have been proposed in ~\cite{low2003natural,liu2020diagnosis,Matsumoto2003Mathematical,wynne2022quantifying}. One of the most popular research topics based on these theoretical derivations is bridge health monitoring by estimating the traffic loads through car-generated vibrations\cite{sekiya2018simplified,obrien2020using,yu2016state,liu2023telecomtm}. A similar idea is also utilized for detecting the location of cracks ~\cite{liu2020damage,nguyen2010multi}. Inspired by these works, ~\cite{bonde2021pignet, dong2023pigsense} leverages the vibration from the pig's movement to monitor the weight gain and other activities of piglets. However, these objects of interest can generate vibrations through themselves. Therefore, these approaches are not adaptable in the store because the item is stationary.

An exterior vibration source is employed for the static item weight estimation, which is referred as active vibration sensing. For example, ~\cite{mirshekari2021obstruction} proves that the footstep-generated vibration can be employed to estimate the weight of the item lying between the footstep and the vibration sensor. ~\cite{codling2021masshog} utilizes a speaker in the pigpen and models a relationship between the pig's weight and its impact on the speaker-generated vibration. However, their systems can only estimate the weight based on the kilogram scale. Vibsense~\cite{liu2017vibsense} utilizes a piezo speaker to generate vibration on a surface actively and identifies the impact of different weights on this vibration signal with a gram scale. VibroScale~\cite{zhang2020vibroscale} turns the smartphone into a weight scale by using the motor and accelerometer in the phone. Nevertheless, none of these systems accounted for the effects of the change of item location on weight estimation. 

To the best of our knowledge, WeVibe represents the first item weight change estimation system using active shelf vibration sensing. WeVibe leverages the physical knowledge from the structural dynamics to characterize the relationship between shelf vibration response and item weight at different locations, which turns out to be linear. WeVibe can quickly adapt to a new location with minimum amount of sensor attached to shelf and data collection effort. 
\section{Conlcusion}
\label{sec:Conlcusion}
We introduce WeVibe, an innovative system that leverages audio-induced shelf vibrations to estimate weight changes in autonomous stores with minimal sensor use and data collection effort. Our approach utilizes structural dynamics to create a scalable, cost-effective solution that outperforms traditional methods by reducing the need for extensive data collection and sensor deployment. We established a physics-informed linear relationship between shelf vibration responses and item weight across different shelf locations through empirical studies and theoretical analysis. This relationship enables WeVibe to quickly adapt to new locations with minimal data, significantly lowering the barrier for deployment in real-world store settings. Our evaluation demonstrates that WeVibe achieved a mean absolute error of up to 48.15g and a standard deviation of 44.96g with one sensor and 10\% samples of three weight classes for each location, which can correctly differentiate weight change bigger than 100g in most cases. In the real item-layout setting, WeVibe can successfully detect if one protein bar is taken or put back at best, confirming the effectiveness of our physics-informed model. We hope our work inspires more adoption of ubiquitous vibration sensing under different structures, not only limited by the gondola shelf.

\bibliographystyle{ACM-Reference-Format}
\bibliography{main}

\end{document}